\documentclass[epjST]{svjour} 

\usepackage[pdftex]{graphicx}

\usepackage{amssymb}
\usepackage{amsmath}
\usepackage{xspace}
\usepackage{color}
\usepackage{dsfont}

\newcommand{\trans}{$\leftrightharpoons$}

\begin{document}

\title{Quantum criticality in dimerised anisotropic spin-1 chains}

\author{Satoshi Ejima\and Florian Lange\and Holger Fehske}
\institute{Institute of Physics, University Greifswald,
17489 Greifswald, Germany}

\abstract{
Applying the (infinite) density-matrix renormalisation group technique, 
we explore the effect of an explicit dimerisation on the ground-state phase diagram of the spin-1 $XXZ$ chain 
with single-ion anisotropy $D$. We demonstrate that the Haldane phase between large-$D$ and 
antiferromagnetic phases survives up to a critical dimerisation only.  As a further new characteristic the dimerisation 
induces a direct continuous Ising quantum phase transition between the large-$D$ and antiferromagnetic phases 
with central charge $c=1/2$, which terminates at a critical end-point where $c=7/10$. Calculating the critical exponents of the 
order parameter, neutral gap and spin-spin-correlation function, we find $\beta=1/8$ (1/24), $\nu=1$ (5/9), and $\eta=1/4$ (3/20),
respectively, which proves the Ising (tricritical Ising) universality class in accordance with field-theoretical predictions.
}

\maketitle

\section{Introduction}
In the last decade, quantum integer-spin chains have received  
revived attention from a topological point of view. In contrast to 
the gapless ground state in the spin-1/2 antiferromagnetic (AFM) Heisenberg chain,
for integer spins there exists a finite gap between the ground state and
the first excited state, as conjectured first by Haldane~\cite{Haldane83}. 
In the spin-1 Heisenberg chain, this so-called Haldane phase is 
a representative of symmetry-protected-topological (SPT) 
phases~\cite{GW09,PTEO10}, which is protected by lattice inversion, 
time-reversal, and dihedral ($\mathbb{Z}_2\times \mathbb{Z}_2$) symmetries. 
The Haldane phase has attracted much attention 
also from an experimental point of view and the Haldane gap was confirmed, 
e.g., in a Ni-compound Ni(C$_2$H$_8$N$_2$)$_2$NO$_2$(ClO$_4$)~\cite{BMAHGH86,RVRERS87}, 
which possesses a small single-ion anisotropy~\cite{DKLM91}.  

A natural extension of the spin-1 $XXZ$ chain with a single-ion anisotropy is an alternating strength of the spin exchange interaction, for example caused by a bond dimerisation,  as realised in a compound
again with Ni$^{2+}$ ions [Ni(333-tet)($\mu$-N$_3$)$_n$](ClO$_4$)$_n$~\cite{HNKKNST98}. 
We expect that such a  bond dimerisation will substantially influence the ground-state properties of the spin model, just as in case of the half-filled extended 
Hubbard model where it opens an energy gap in the spin-density-wave regime and leads to new types of quantum phase transitions~\cite{EELF16}.

At present the most reliable method for an exact numerical treatment of one-dimensional (strongly correlated) electron, boson and spin systems 
seems to be the density-matrix renormalization group (DMRG) technique, which is based on a matrix-product state (MPS) approximation of the ground state~\cite{White92,Sch11}. 
Here, we employ an infinite DMRG (iDMRG) variant that uses an infinite MPS  representation 
and thus---working in the thermodynamic limit---avoids any boundary and finite-size effects~\cite{Mc08}.

\section{Model and ground-state phase diagram}
\begin{figure}[h!]
 \centering
 \includegraphics[width=0.95\columnwidth,clip]{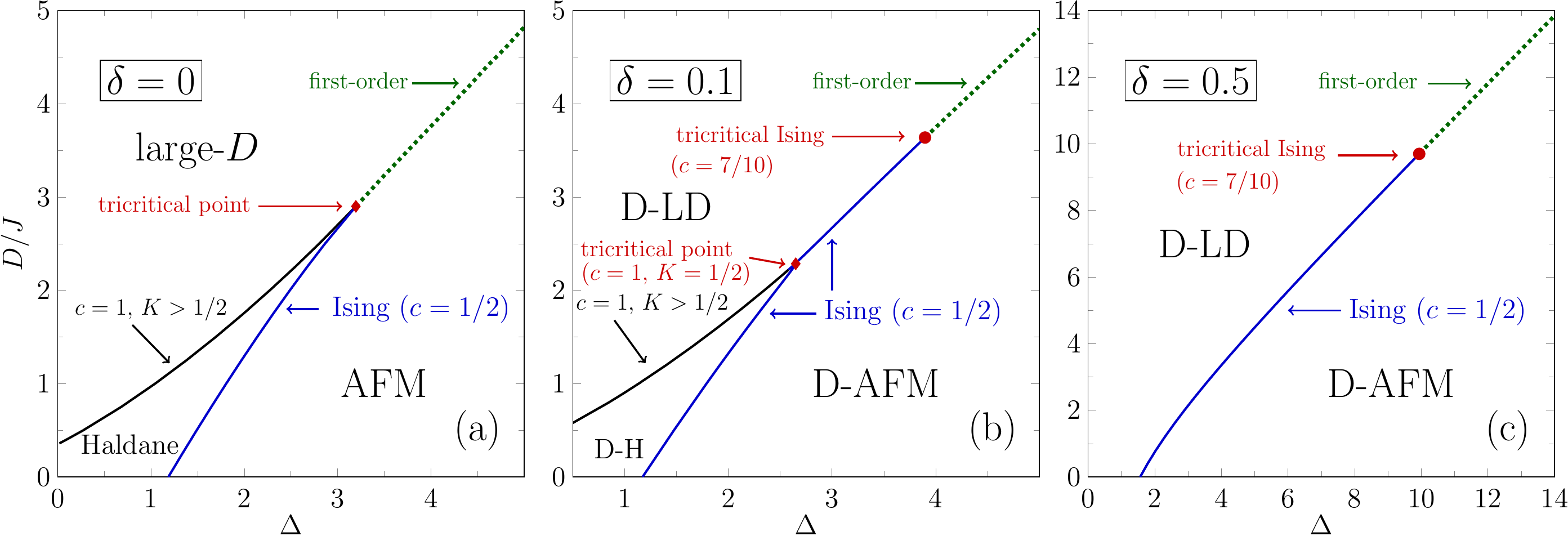}
 \caption{IDMRG ground-state phase diagram of the spin-1 $XXZ$ 
 chain~\eqref{hamil} with dimerisation  $\delta=0$ (a), $0.1$ (b) and $0.5$ (c). The phase boundaries are determined by means of the methods described in our previous work~\cite{EF15,Scipost18}.
 }
 \label{pd}
\end{figure}

The Hamiltonian of a dimerised spin-1 AFM spin chain with single-ion anisotropy $D$ reads
\begin{eqnarray}
 \hat{H}=J\sum_j(1+\delta(-1)^j)(\hat{\bf S}_j\cdot\hat{\bf S}_{j+1})_\Delta+D\sum_j(\hat{S}_j^z)^2\,,
 \label{hamil}
\end{eqnarray}
where $(\hat{\bf S}_j\cdot\hat{\bf S}_{j+1})_\Delta
      =\hat{S}_j^x\hat{S}_{j+1}^x+\hat{S}_j^y\hat{S}_{j+1}^y
       +\Delta \hat{S}_j^z\hat{S}_{j+1}^z$.
For vanishing dimerisation $\delta$ and not too large $\Delta$ and $D/J$,  the ground-state phase
diagram of the model~(\ref{hamil}) develops a nontrivial topological Haldane phase between topologically trivial large-$D$ (LD)
and AFM phases [see Fig.~\ref{pd}(a)]. 
The phase transitions from Haldane to the LD and AFM phases 
belong to the Gaussian and Ising universality classes, respectively, 
while the LD{\trans}AFM transition is of first order.   

A finite dimerisation strongly affects the phase boundary between the dimerised LD (D-LD) and AFM (D-AFM)
phases: Now, at sufficiently large $\Delta$ and $D/J$, a new direct Ising transition appears with $c=1/2$, which terminates at a tricritical Ising point
where $c=7/10$~\cite{Scipost18}. Beyond this critical end-point, in the very strong-coupling regime,
the quantum phase transition becomes first order [see Fig.~\ref{pd}(b)].
Increasing dimerisation further, the dimerised Haldane phase (D-H)
disappears at about $\delta\gtrsim0.26$, if we limit ourselves 
to the case $J,D>0$. Figure~\ref{pd}(c) shows a situation, where   
only D-LD and D-AFM phases appear, for $\delta=0.5$. 

\section{Critical exponents}
We now provide further evidence for the Ising and  tricritical 
Ising criticality at the  D-LD$\leftrightharpoons$D-AFM quantum phase transition 
of the dimerised $XXZ$ chain~\eqref{hamil}. Approaching a continuous quantum phase transition, the system behaviour is characterised by a set of universal exponents describing the power-law dependencies of the relevant order parameter, correlation functions or excitation gaps on $(g-g_c)$, where  $g$ parametrizes the (general) coupling constant of the model under consideration and $g_c$ is its critical value at the transition point.  Assuming the Ising (tricritical Ising) universality class, 
the critical exponents of the order parameter, neutral gap and spin-spin-correlation function should be $\beta=1/8$ (1/24), $\nu=1$ (5/9), and $\eta=1/4$ (3/20)~\cite{FMS97}, and will be linked with each other by the scaling relation $\tfrac{\nu}{2}(\eta+d-2)=\beta$, where $d$ is the spatial dimension. In what follows, this will be proved for the rather complex model~\eqref{hamil} performing unbiased iDMRG simulations. 
Note that it is quite challenging to determine  the critical exponent, especially in the latter case, because the critical end-point itself has to be determined first with high precision by varying $\Delta$ and $D$ simultaneously.

\subsection{Order parameter}
The exponent $\beta$ can be extracted from the AFM order parameter 
\begin{eqnarray}
 \langle\hat{m}_{\rm AFM}\rangle=\frac{1}{L}\sum_j(-1)^j\langle\hat{S}_j^z\rangle\,.
 \label{eq:op}
\end{eqnarray}

\begin{figure}[b!]
 \centering
 \includegraphics[width=0.9\textwidth,clip]{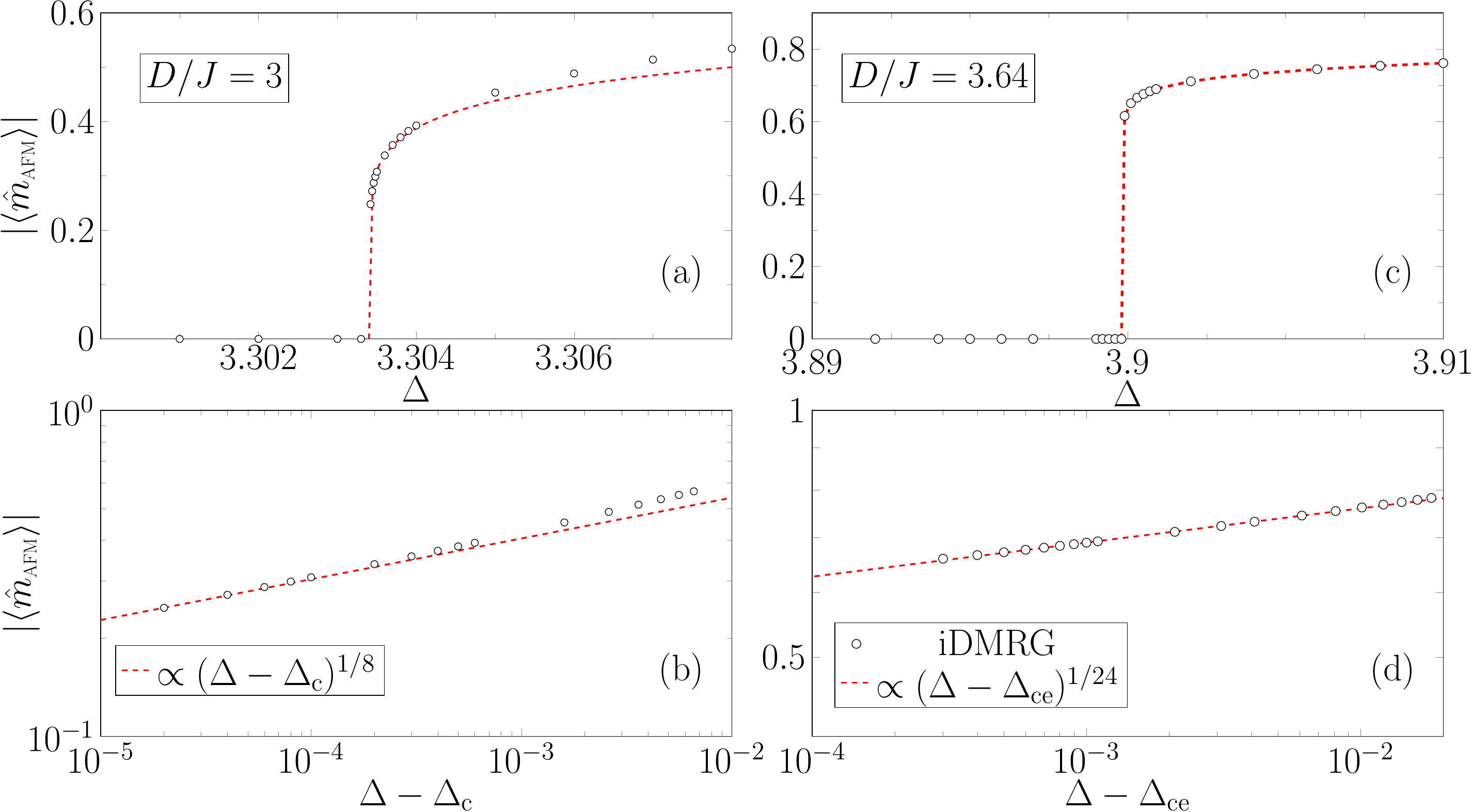}
 \caption{
Absolute value of the AFM order parameter  $|\langle\hat{m}_{\rm AFM}\rangle|$ 
 (upper row) in the vicinity of the D-LD$\leftrightharpoons$D-AFM Ising (a) and tricritical Ising (c) transition for $\delta=0.1$ at
 fixed $D/J=3$ and $D/J=3.64$, respectively. Symbols are iDMRG data obtained with bond dimension $\chi=800$, dashed lines display 
 the fitting function with critical exponents $\beta=1/8$ (a) and 1/24 (c). Here the relative error of the exponents is less than 7\% and might be reduced further increasing the bond dimension $\chi$. The corresponding log-log plots (b) and (d) in the lower row demonstrate that these power-laws apply with high precision close to the critical points.
 Here, $\Delta_{\rm c}\simeq 3.303$ and $\Delta_{\rm ce}\simeq 3.900$.}
 \label{beta}
\end{figure}

Figure~\ref{beta} displays $|\langle\hat{m}_{\rm AFM}\rangle|$ across the Ising and tricritical Ising D-LD$\leftrightharpoons$D-AFM  transitions when the spin anisotropy $\Delta$ is increased at fixed $D/J=3$ and $D/J=3.64$, respectively. The corresponding critical couplings are $\Delta_{\rm c} = 3.303$ and $\Delta_{\rm ce} = 3.900$. Evidently, the onset of $|\langle \hat{m}_{\rm AFM} \rangle |$ in the D-AFM phase is much more abrupt for the transition at the critical end point. 
More importantly the critical exponents of the order parameter function are verified to be 1/8 and 1/24 at the Ising transition with $c=1/2$ and  the tricritical Ising transition with $c=7/10$, respectively. 

\subsection{Neutral gap}
\begin{figure}[h!]
 \centering
 \includegraphics[width=0.9\textwidth,clip]{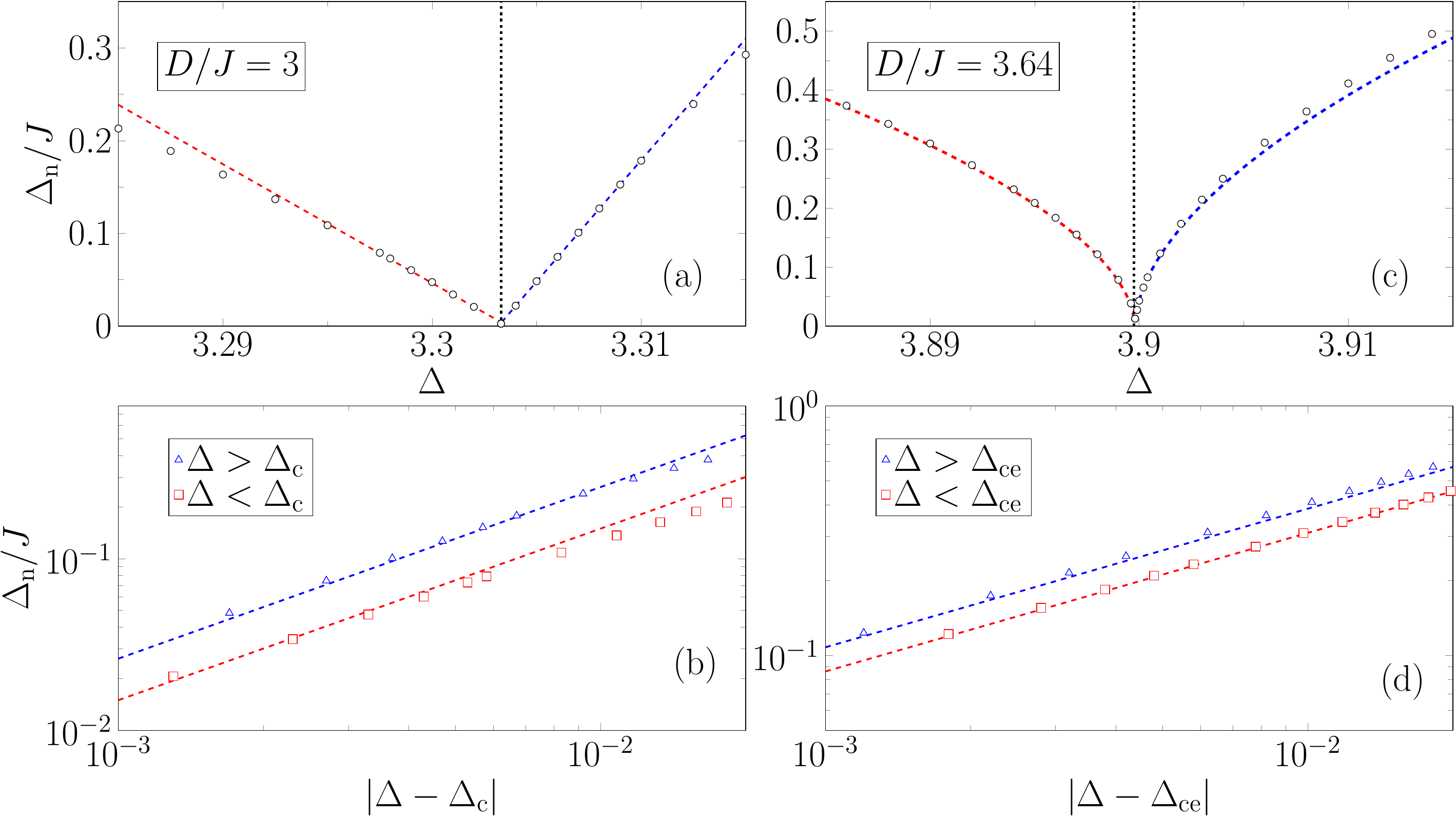}
 \caption{Closing of the neutral gap (upper row) at the Ising (a) and tricritical Ising (c) transition for the same parameters as in Fig.~2. 
 The log-log plots (lower row) show that the $L\to\infty$ extrapolated DMRG data 
  can be excellently fitted by $|\Delta-\Delta_{\rm c}|^\nu$ with $\nu=1$ (b) and  
 $|\Delta-\Delta_{\rm ce}|^\nu$ with $\nu=5/9$ (d).
 }
 \label{nu}
\end{figure}

The neutral gap is defined as
\begin{eqnarray}
 \Delta_{\rm n}(L)=E_1-E_0\,,
 \label{eq:nu}
\end{eqnarray}
where $E_0$ ($E_1$) is the energy of the ground state (first excited state) for a system with $L$ sites and 
vanishing total spin $z$-component, which is directly accessible by DMRG. To determine the neutral gap we use a DMRG representation 
with infinite boundary conditions~\cite{PVM12}. Figures~\ref{nu}~(a) and~\ref{nu}~(b)  show  that the neutral gap closes and opens linearly on passing the Ising transition, i.e., $\nu=1$. Also for the tricritical Ising point the prediction of perturbed conformal field theory could be confirmed: Our data clearly yield $\nu=5/9$, see Figs.~\ref{nu}~(c) and~\ref{nu}~(d).  

\subsection{Spin-spin correlation function}
Let us finally investigate the critical behaviour of the staggered $z$-$z$ spin correlator 
$\langle\hat{n}^z_j\hat{n}^z_{j+\ell}\rangle$
with $\hat{n}^z_j=(-1)^j (\hat{S}^z_j-\hat{S}^z_{j+1})/2$. For this we determine the exponent $\eta$  
from the spatial decay of the correlations at large distances  $\ell \gg 1$:
\begin{eqnarray}
 \langle\hat{n}^z_j\hat{n}^z_{j+\ell}\rangle 
   \propto \ell^{-\eta}\,. 
 \label{eq:ss}
\end{eqnarray}
\begin{figure}[h!]
 \centering
 \includegraphics[width=0.9\textwidth,clip]{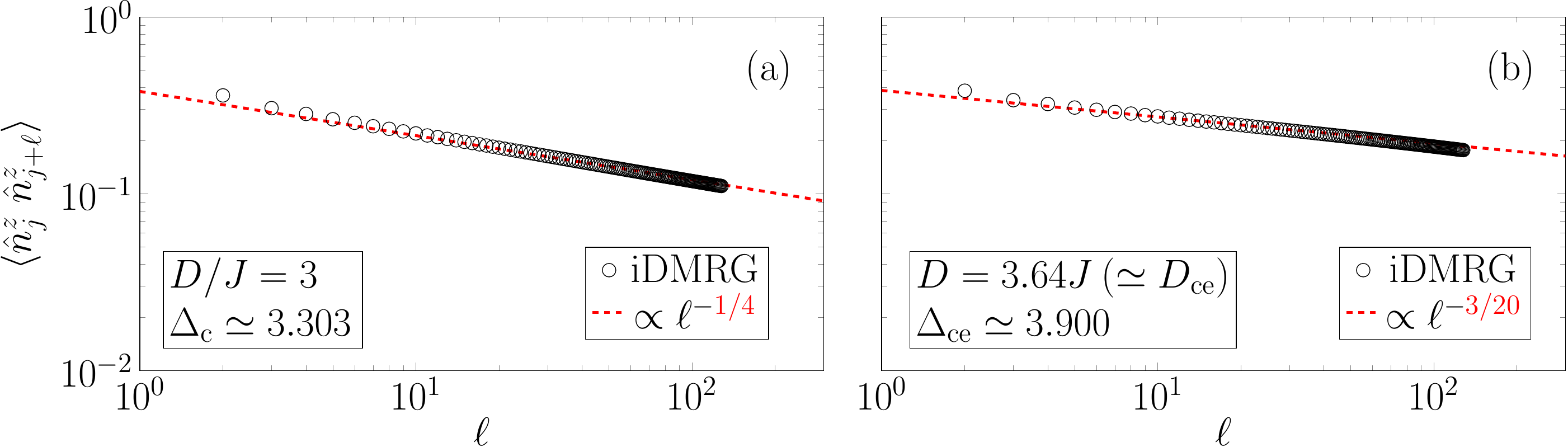}
 \caption{Decay of the longitudinal spin-spin two-point functions at Ising (a) 
 and tricritical Ising (b) transition points for $\delta=0.1$.
 Symbols give iDMRG data; 
 power-laws predicted by field theory drawn in  by dotted lines.
 }
 \label{eta}
\end{figure}
Figure~\ref{eta} clearly shows that the critical exponents obtained in such a way from the iDMRG results are $\eta=1/4$ and $3/20$.  That means, according to field theory, the transitions belong to the Ising respectively  tricritical Ising universality class.

\section{Summary}
To conclude, we have examined numerically the criticality of the quantum phase transition between the dimerised large-$D$ and antiferromagnetic phases in the spin-1 $XXZ$ chain with explicit bond dimerisation and proved the Ising ($c=1/2$) and tricritical Ising ($c=7/10$) universality classes. The critical exponents extracted from large-scale (infinite matrix-product-state based) density-matrix renormalisation group  simulations corroborate the predictions of bosonisation-based field theory, and consequently fulfil the desired scaling relation. The calculated ground-state phase  diagram shows that a symmetry-protected topological Haldane phase appears---in between large-$D$ and antiferromagnetic phases---for not too large bond dimerisation only.

We note that a  similar situation evolves in the one-dimensional extended Bose-Hubbard model 
with bond dimerisation~\cite{Sugimoto}.

\begin{acknowledgement}
 The DMRG simulations were performed using the ITensor library~\cite{ITensor}.   
 S.E. and F.L. were supported by Deutsche Forschungsgemeinschaft through project EJ 7/2-1  
 and FE 398-1, respectively.
 We thank F.H.L. Essler, Y. Ohta and T. Yamaguchi for valuable discussions.  
\end{acknowledgement}


\end{document}